\documentclass[12pt]{article}
\usepackage{xspace}
\usepackage{epsfig}
\usepackage{cite}
\newcommand{\qsq    }{\ensuremath{Q^{2}}\xspace}
\newcommand{\der    }{\ensuremath{\mathrm{d}}\xspace}
\newcommand{\kt     }{\ensuremath{k_{\mathrm{t}}}\xspace}
\newcommand{\pT     }{\ensuremath{p_{\mathrm{T}}}\xspace}
\newcommand{\gev    }{\ensuremath{\rm GeV}\xspace}
\newcommand{\gevsq  }{\ensuremath{\rm GeV^2}\xspace}

\newcommand{\epem   }{\ensuremath{\rm e^+e^-}\xspace}
\newcommand{\emi    }{\ensuremath{\rm e^-}\xspace}

\newcommand{\ep     }{\ensuremath{\rm e\rm p}}\xspace
\xspace
\newcommand{\ggss   }{\ensuremath{\gamma^{\star}\gamma^{\star}}\xspace}
\newcommand{\eb     }{\ensuremath{E_{\rm b}}\xspace}
\newcommand{\stt    }{\ensuremath{\sigma_\mathrm{TT}}\xspace}
\newcommand{\slt    }{\ensuremath{\sigma_\mathrm{LT}}\xspace}
\newcommand{\stl    }{\ensuremath{\sigma_\mathrm{TL}}\xspace}
\newcommand{\sll    }{\ensuremath{\sigma_\mathrm{LL}}\xspace}
\newcommand{\ttt    }{\ensuremath{\tau_\mathrm{TT}}\xspace}
\newcommand{\ttl    }{\ensuremath{\tau_\mathrm{TL}}\xspace}
\newcommand{\LTT    }{\ensuremath{L_\mathrm{TT}}\xspace}
\newcommand{\calLTT }{\ensuremath{{\cal L}_\mathrm{TT}}\xspace}
\newcommand{\sgg    }{\ensuremath{\sigma_{\gs\gs}}\xspace}
\newcommand{\barph  }{\ensuremath{\bar{\phi}}\xspace}
\newcommand{\cosph  }{\ensuremath{\cos\barph}\xspace}
\newcommand{\costph }{\ensuremath{\cos 2\barph}\xspace}
\newcommand{\mupmum }{\ensuremath{\mu^+\mu^-}\xspace}
\newcommand{\lplm   }{\ensuremath{\ell^+\ell^-}\xspace}
\newcommand{\wsq    }{\ensuremath{W^{2}}\xspace}
\newcommand{\see    }{\ensuremath{s_{\rm ee}}\xspace}
\newcommand{\ssee   }{\ensuremath{\sqrt{\see}}\xspace}
\newcommand{\gs     }{\ensuremath{\gamma^{\star}}\xspace}
\newcommand{\ppbar  }{\ensuremath{\mathrm{p\bar{p}}}\xspace}
\newcommand{\pp     }{\ensuremath{\mathrm{p p}}\xspace} 
\newcommand{\gp     }{\ensuremath{\mathrm{\gamma p}}\xspace} 
\newcommand{\Ybar   }{\ensuremath{\overline{Y}}\xspace} 
\newcommand{\al     }{\ensuremath{\alpha_s}\xspace} 
\newcommand{\aem    }{\ensuremath{\alpha}\xspace} 
\newcommand{\lamf   }{\ensuremath{\Lambda_{\rm 5}^{\scriptstyle%
                      \overline{\rm MS}}}\xspace}
\newcommand{\eeqq   }{\ensuremath{\epem\rightarrow\epem\,q\bar{q}}\xspace} 
\begin{document}
\begin{titlepage}
\begin{center}
{\Large  EUROPEAN ORGANIZATION FOR NUCLEAR RESEARCH}
\end{center}
\bigskip\bigskip
\begin{flushright}
       CERN-EP-2001-064 \\
       31 August 2001
\end{flushright}
\begin{center}
{\huge\bf\boldmath
 Measurement of the Hadronic \\
 Cross-Section for the Scattering of \\\vspace{0.25cm}
 Two Virtual Photons at LEP 
}\end{center}
\bigskip
\bigskip
\begin{center}{\LARGE The OPAL Collaboration}\end{center}
\bigskip\bigskip
\bigskip
\begin{center}{\large  Abstract}\end{center}
 The interaction of virtual photons is investigated
 using the reaction $\epem \rightarrow \epem\,\mbox{hadrons}$ based
 on data taken by the OPAL experiment at \epem centre-of-mass 
 energies $\ssee=189-209$~\gev, for $W>5$~GeV and at an average 
 \qsq of 17.9~\gevsq.
 The measured cross-sections are compared to predictions of the 
 Quark Parton Model (QPM), to the Leading Order QCD Monte Carlo
 model PHOJET to the NLO prediction for the reaction \eeqq, 
 and to BFKL calculations.
 PHOJET, NLO \eeqq, and QPM describe the data reasonably well, whereas 
 the cross-section predicted by a Leading Order BFKL calculation is 
 too large.
\mbox{ }\\
\bigskip\bigskip\bigskip
\begin{center} {\large (Submitted to European Physical Journal C)}
\end{center}
\end{titlepage}
%
%
\begin{center}{\Large        The OPAL Collaboration
}\end{center}\bigskip
\begin{center}{
G.\thinspace Abbiendi$^{  2}$,
C.\thinspace Ainsley$^{  5}$,
P.F.\thinspace {\AA}kesson$^{  3}$,
G.\thinspace Alexander$^{ 22}$,
J.\thinspace Allison$^{ 16}$,
G.\thinspace Anagnostou$^{  1}$,
K.J.\thinspace Anderson$^{  9}$,
S.\thinspace Arcelli$^{ 17}$,
S.\thinspace Asai$^{ 23}$,
D.\thinspace Axen$^{ 27}$
% ...
}\end{center}\bigskip
\bigskip
$^{  1}$School of Physics and Astronomy, University of Birmingham,
Birmingham B15 2TT, UK
\newline
$^{  2}$Dipartimento di Fisica dell' Universit\`a di Bologna and INFN,
I-40126 Bologna, Italy
\newline
$^{  3}$Physikalisches Institut, Universit\"at Bonn,
D-53115 Bonn, Germany
\newline
% ...
$^{  5}$Cavendish Laboratory, Cambridge CB3 0HE, UK
\newline
% ...
$^{  9}$Enrico Fermi Institute and Department of Physics,
University of Chicago, Chicago IL 60637, USA
\newline
% ...
$^{ 16}$Department of Physics, Schuster Laboratory, The University,
Manchester M13 9PL, UK
\newline
$^{ 17}$Department of Physics, University of Maryland,
College Park, MD 20742, USA
\newline
% ...
$^{ 22}$Department of Physics and Astronomy, Tel Aviv University,
Tel Aviv 69978, Israel
\newline
$^{ 23}$International Centre for Elementary Particle Physics and
Department of Physics, University of Tokyo, Tokyo 113-0033, and
Kobe University, Kobe 657-8501, Japan
\newline
% ...
$^{ 27}$University of British Columbia, Department of Physics,
Vancouver BC V6T 1Z1, Canada
\newline
% ...
Tel Aviv 69978, Israel.
%
%
\section{Introduction}
\label{sec:intro}
 The classical way to investigate the structure of the photon at
 \epem colliders is the measurement of the process
%
\begin{equation}
 \mathrm{e^+}(p_1) \mathrm{e^-}(p_2) \rightarrow \mathrm{e^+}(p_1^\prime) 
 \mathrm{e^-}(p_2^\prime)\,\mbox{X},
 \label{eqn:react}
\end{equation}
%
 proceeding via the interaction of two photons, which can be either 
 quasi-real, $\gamma$, or virtual, \gs.
 The terms in brackets represent the four-vectors of the particles 
 as shown in Fig.~\ref{feyn_diag}.
 \par
 Depending on the virtualities of the photons the scattered
 electrons\footnote{Electrons and positrons are generically
 referred to as
 electrons.} may be observed in the detector.
 In the case where none of the electrons is observed
 (anti-tagged), the structure of the 
 quasi-real photon has been studied by OPAL in terms of total 
 cross-sections~\cite{OPALSTOT}, jet production~\cite{OPALPR250}, and
 charm production~\cite{OPALPR294}.
 If only one electron is observed (single-tagged), the process can be 
 described as deep-inelastic electron scattering off a quasi-real photon.
 These events have been studied by OPAL to measure the QED and QCD
 photon structure functions~\cite{OPALPR271,OPALPR294,OPALPRSTAG,OPALPR314}.
 If both electrons are observed (double-tagged), the dynamics of
 highly virtual photon collisions is probed.
 The QED structure of the interactions of two highly virtual photons has
 already been studied by OPAL~\cite{OPALPR271}. 
 In the analysis presented here, the investigation 
 is extended to the measurement of the hadronic structure.
 The results are based on data recorded by the OPAL experiment at LEP
 in the years 1998 to 2000
 at \epem centre-of-mass energies of $\ssee=189-209$~\gev,
 using events where both scattered electrons are observed in the
 small-angle silicon-tungsten (SW) luminometer.
 The measured differential cross-sections are compared to the prediction of
 the Quark Parton Model (QPM), to a NLO calculation~\cite{Cacciari} of the
 process \eeqq, to the PHOJET Monte Carlo model~\cite{PHOJET1.10} 
 and to BFKL~\cite{BFKL} calculations~\cite{carlo,kim,kwiecinski}. 
 A similar analysis has been published by the L3 Collaboration~\cite{L3_paper}
 using data taken at $\ssee=91$~\gev and $\ssee=183$~\gev.
%
%
\section{Theoretical framework}
%
\begin{figure}[t]
\begin{center}
{\includegraphics[width=0.6\linewidth]{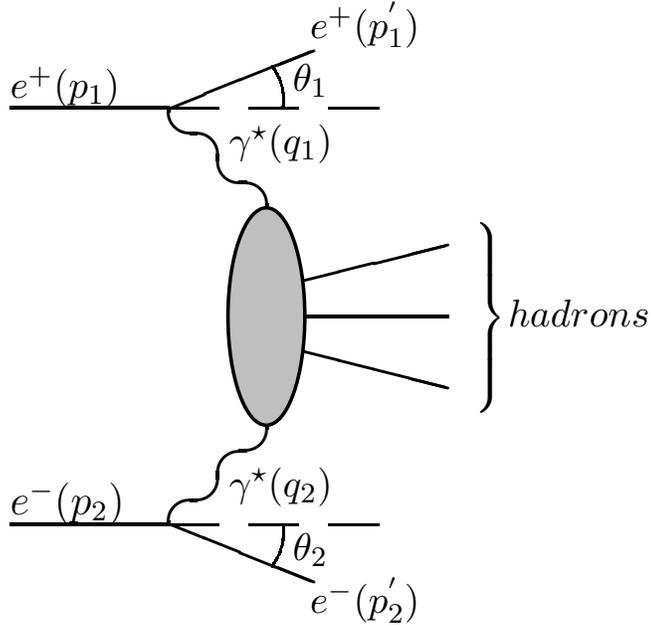}}
\caption{The diagram corresponding to the process 
         $\epem \rightarrow \epem\,\mbox{hadrons}$.}
\label{feyn_diag}
\end{center}
\end{figure}
%
 In this paper double-tagged events are studied, i.e.~both final 
 state electrons are scattered at sufficiently large polar 
 angles\footnote{The right-handed OPAL coordinate system is defined
 with the $z$ axis pointing in the direction of the $\emi$ beam and
 the $x$ axis pointing towards the centre of the LEP ring. The polar
 angle $\theta$, the azimuthal angle $\phi$ and the radius $r$ are the usual
 spherical coordinates.} $\theta_{i}$ to be observed in the detector. 
 This corresponds to the situation where both radiated photons which take
 part in the hard scattering process, are highly virtual. 
 Throughout the paper, $i=1,2$ denotes quantities which are connected with
 the upper and lower vertex in Fig.~\ref{feyn_diag}, respectively.
 \par
 The virtualities of the radiated photons are given by
 $Q_{i}^{2}=-(p_{i}-p_{i}^{'})^{2}>0$. 
 The usual dimensionless variables of deep inelastic scattering are
 defined as:
 \begin{equation}
 y_{1}=\frac{q_{1}q_{2}}{p_{1}q_{2}}, \;\;\;\;\;\;\;\;\;\; 
 y_{2}=\frac{q_{1}q_{2}}{p_{2}q_{1}},
 \end{equation}
 \begin{equation}
 x_{1}=\frac{Q_{1}^{2}}{2q_{1}q_{2}}, \;\;\;\;\;\;\;\;\;
 x_{2}=\frac{Q_{2}^{2}}{2q_{1}q_{2}}.
 \end{equation}
 The $\epem$ centre-of-mass energy squared is given by $\see=(p_{1}+p_{2})^{2}$
 and the hadronic invariant mass squared by $W^{2}=(q_{1}+q_{2})^2$.

 The kinematical variables $Q_{i}^{2}$, $y_{i}$ and $x_{i}$ are obtained
 from the four-vectors of the tagged electrons and the hadronic final
 state via:
 \begin{eqnarray}
 Q_{i}^{2} & = & 2E_{\rm b}E_{i}(1-\cos\theta_{i}), \\
 y_{i} & = & 1-\frac{E_{i}}{E_{\rm b}}\cos^{2}(\theta_{i}/2), \\
 x_{i} & = & \frac{Q^{2}_{i}}{Q^{2}_{1}+Q^{2}_{2}+W^{2}},
 \end{eqnarray}
 where $E_{\rm b}$ refers to the energy of the beam electrons,
 and the mass $m_{\mathrm e}$ of the electron has been neglected. 
 \par
 In this analysis, the hadronic invariant mass, $W$, is obtained
 from the energies, $E_{h}$, and momenta, $\vec{p}_{h}$, of final state
 hadrons ($h$), excluding the scattered electrons:
%
 \begin{equation}
 W^{2}=\left(\sum_{h}E_{h}\right)^{2}-\left(\sum_{h}\vec{p}_{h}\right)^{2}
      = E_{\rm had}^2 - {\vec{p}_{\rm had}}^{\,\,2}\,.
 \end{equation}
%
 \par
 For the comparison of the data to BFKL calculations the following 
 additional kinematic quantity, which is a measure of the 
 length of the gluon ladder, is defined~\cite{bartels}:
%
 \begin{equation}
 Y      = \ln\left(\frac{\see y_{1}y_{2}}{\sqrt{Q^{2}_{1}Q^{2}_{2}}}\right) 
   \simeq  \ln\left(\frac{\wsq}{\sqrt{Q^{2}_{1}Q^{2}_{2}}}\right) 
        = \Ybar\, ,
 \label{eqn:ybar}
 \end{equation}
%
 where the approximation requires $\wsq\gg Q^{2}_{i}$.
 In the analysis presented here $Y$ will be approximated by \Ybar.
 \par
 The differential cross-section for the process of Eq.~\ref{eqn:react}
 in the limit $Q_{i}^{2}\gg m_{\mathrm e}^2$ and for small
 values of $y_{i}$ is given by~\cite{BUD-7501}
%
 \begin{eqnarray}
 \mathrm{d}^6\sigma 
  &=&
  \frac{\mathrm{d^3p}_1^\prime\mathrm{d^3p}_2^\prime}{E_1^\prime\,E_2^\prime}
  \,\calLTT
  \left(
  \stt + \slt + \stl + \sll + \frac{1}{2}\ttt\costph - 4\ttl\cosph
  \right)  \label{eqn:truesimp}
  \nonumber\\
  &=&
  \frac{\mathrm{d^3p}_1^\prime\mathrm{d^3p}_2^\prime}{E_1^\prime\,E_2^\prime}
  \,\calLTT\,\sgg\,,\\
  \LTT  &=& \int \frac{\mathrm{d^3p}_1^\prime\mathrm{d^3p}_2^\prime}
 {E_1^\prime\,E_2^\prime}\calLTT,
 \end{eqnarray}
%
 where \barph is the angle between the two scattering planes 
 of the electrons in the photon-photon centre-of-mass system.
 The cross-sections \stt, \stl, \slt and \sll and the interference terms
 \ttt and \ttl correspond to specific helicity states of the interacting
 photons (T=transverse and L=longitudinal).
 The sum of these terms is \sgg, the cross-section
 for the reaction $\ggss \rightarrow\,\mbox{hadrons}$.
 The process $\epem \rightarrow \epem\,\mbox{hadrons}$ can be written as 
 a product of a term describing the flux of the incoming photons and 
 the cross-section for the interaction of the virtual photons.
 For any bin \LTT is derived by integrating over the phase space of the
 outgoing electrons.
 Here \LTT describes the flux of transversely polarised photons, and only
 depends on the four-vectors $q_{1}$, $q_{2}$, $p_1$, $p_2$
 and on the mass of the electron~\cite{bib-Nisius}.
 \par
 In the case of lepton pair production $\epem \rightarrow \epem\,\lplm$
 the cross-section is completely determined by QED. 
 For hadronic final states QCD corrections have to be taken
 into account and in addition
 the cross-sections and interference terms cannot be 
 completely calculated within the framework of perturbative QCD.
 For \mupmum final states~\cite{OPALPR271} it is found that
 the contributions from \ttt and \ttl are large for
 $Q_{1}^{2} \approx Q_{2}^{2}$ and  small values of \wsq.
 It cannot be excluded that the interference terms and 
 contributions from longitudinal photons in the QCD case are also
 large~\cite{bib-Nisius}. 
 Therefore, only comparisons to theoretical models containing predictions 
 for all cross-sections and interference terms are meaningful.
 The cleanest experimental quantity which can be extracted without
 making further assumptions about the interference terms
 is the cross-section for the reaction 
 $\epem \rightarrow \epem \,\mbox{hadrons}$ as given in Eq.~\ref{eqn:truesimp}.
 Dividing by \LTT the cross-section \sgg can be extracted.
 \par
 The order \al QCD radiative corrections to the process \eeqq have
 also been calculated~\cite{Cacciari}.
 The calculation accounts for the running of \al and \aem and allows
 for the evaluation of differential cross-sections for several variables, 
 but, it is valid only for massless quarks.
 The LO result is somewhat higher than the QPM prediction, which is expected 
 due to the use of massless quarks.
 Overall the NLO corrections are predicted to be small, but they change 
 e.g.~the shape of the \Ybar distribution, as can be seen from 
 Table~\ref{t_nlo}.
 \par
 Recently, much attention has been given to the BFKL pomeron~\cite{BFKL}, 
 especially for small Bjorken-$x$
 deep-inelastic electron-proton ($\ep$) scattering at HERA. 
 With increasing centre-of-mass energies squared, $s$, and for moderate
 photon virtualities, large logarithms in $1/x$ are expected to affect 
 the rise of the cross-section. 
 Resummation of these logarithms leads to the so-called 
 BFKL evolution equation from which one derives
 that the total ep cross-section should increase as $s^{\alpha_0}$
 with $\alpha_0 \sim 0.5$ in Leading Order (LO), where $\alpha_0$ denotes the
 pomeron intercept. The onset of such BFKL
 effects has been searched for in ep structure function and hadronic
 final state data~\cite{BFKL-HERA}. The situation is so far still
 inconclusive, with the strongest hint for BFKL effects in the data 
 coming from the analysis of forward $\pi^0$ production.
 \par
 It has been argued~\cite{bartels,brodsky,bialas,boonenkamp} that
 \epem collider offer an excellent opportunity to test the BFKL
 prediction, through a measurement of \sgg.
 For sufficiently large photon virtualities $Q_{1}^{2}$ and $Q_{2}^{2}$
 (i.e.more than a few GeV$^2$), 
 this BFKL calculation can be carried out without 
 non-perturbative input. If additionally $Q_{1}^{2} \simeq Q_{2}^{2}$,
 then the evolution in $Q^2$ is suppressed, allowing for a clean test
 of BFKL effects. This condition is to a good approximation fulfilled
 for the data sample selected in this paper.
 \par
\begin{figure}[t]
\begin{center}
{\includegraphics[width=1.0\linewidth]{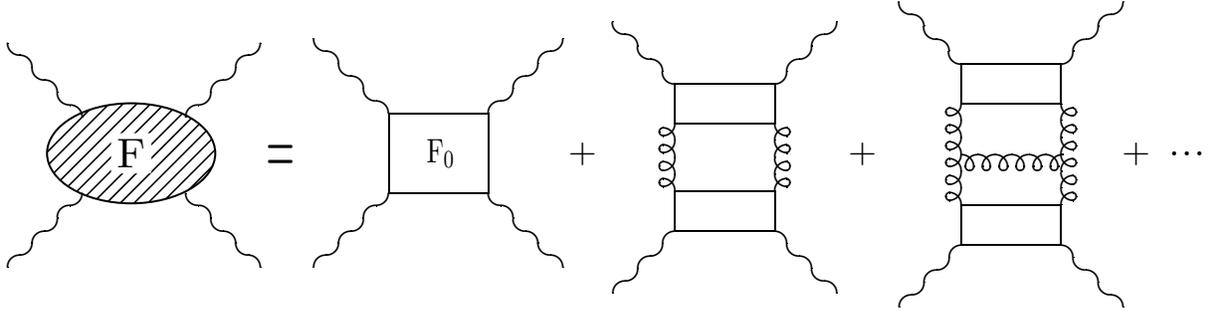}}
\caption{Diagrams contributing to the leading $\ln1/x$ approximation for the 
         \ggss cross-section.}
\label{tg_sum}
\end{center}
\end{figure}

 The leading-log resummation for $\ggss$ scattering, can be represented
 by the so-called `gluon ladder diagrams', as sketched in Fig.~\ref{tg_sum}.
 The first diagram is the $\gamma\gamma \rightarrow q\bar{q}$
 contribution and the next ones give the gluon exchange contributions.
 The original LO-BFKL calculations~\cite{bartels} predicted an increase of 
 \sgg by a factor 20 or more compared to calculations 
 without BFKL effects, or with only DGLAP $(Q^2)$ evolution~\cite{DGLAP}. 
 Since then the LO-BFKL
 calculations have been improved by including charm quark mass effects, 
 running of the strong coupling constant $\alpha_s$ and the 
 contribution of the longitudinal photon polarisation states. Recently,
 it has become clear that
 the Next-to-Leading Order (NLO) corrections to the BFKL equation
 are large and effectively reduce 
 the value of $\alpha_0$. 
 A phenomenological determination of the Higher Order (HO) effects
 was presented in Ref.~\cite{boonenkamp} and the 
 resulting BFKL scattering cross-sections 
 were shown to increase by a factor 2-3 only, relative to the calculations
 without BFKL effects.
 Since then theoretically motivated improved higher order 
 calculations have been 
 performed~\cite{kim,kwiecinski,ciafaloni,schmidt,forshaw,thorne,Altarelli},
 and give similar results.
%
%
 \section{Monte Carlo generators}
 Several Monte Carlo generators have been compared to the data 
 or have been used to correct the data for detector effects.
 Most relevant to the analysis presented here are the programs which are 
 used to model two-photon interactions for double-tagged, single-tagged and 
 anti-tagged events. 
 The main features of the programs are briefly described here.
 For further details the reader is referred to the original publications. 
 An overview can be found in~\cite{bib-Nisius}.
 \par
 The PHOJET event generator has been used to simulate double-tagged
 events. There are significant differences between PHOJET1.10 and 
 the previous version PHOJET1.05 in several aspects of the event generation,
 the latter producing e.g. larger cross-sections at large $W$. In this paper 
 we compare with the most recent and improved version of the program 
 PHOJET1.10.
 The program is based on the Dual Parton Model, containing both hard and soft 
 processes~\cite{PHOJET1.10}. 
 The hard processes are calculated in LO perturbative QCD, and
 soft processes are modelled based on $\gp$, $\pp$ and $\ppbar$ data assuming 
 Regge factorisation.
 The \ggss cross-section is obtained from the $\gamma\gamma$ 
 cross-section by extrapolating in \qsq on the basis of the Generalised 
 Vector Dominance model using the description of Ref.~\cite{bib-ginz}.
 Events are generated for both soft and hard partonic processes. A
 cut-off on the transverse momentum of the scattered partons in the 
 photon-photon centre-of-mass system of 2.5~GeV is used to separate the
 two classes of events.
 For this reason the generation of events with $W$ below 5~GeV is known to
 be incomplete.
 As a cross-check of the PHOJET model the PYTHIA6.130~\cite{bib-pythia} 
 Monte Carlo generator,  based on a recent model by Friberg and 
 Sj\"ostrand~\cite{bib-Chris}, is used for the simulation of double-tagged
 two-photon events.
 \par
 The general purpose Monte Carlo program HERWIG5.9+\kt(dyn)~\cite{bib-HERWIG} 
 was used to simulate single-tagged two-photon events.
 This model version uses a modified transverse momentum
 distribution, \kt, for the quarks inside the photon, with the upper
 limit dynamically (dyn) adjusted according
 to the hardest scale in the event, which is of order \qsq.
 This has been found~\cite{OPALPR316} to give a better description of the
 experimentally observed hadronic final states.
 \par
 The PHOJET1.10 Monte Carlo has also been used to simulate anti-tagged 
 two-photon events. It is known to  describe satisfactorily
 the OPAL anti-tagged two-photon data~\cite{OPALSTOT}.
 \par
 The QPM cross-section $\epem \rightarrow \epem q\overline{q}$,
 which corresponds to the diagram labelled $F_0$ in Fig.~\ref{tg_sum},
 was calculated with the GALUGA~\cite{GALUGA} program, which includes all
 terms from Eq.~\ref{eqn:truesimp}.
 The quark masses assumed are 0.325~GeV for uds and 1.5~GeV for c quarks. 
 For the region of $W>5$~\gev considered here, the cross-section 
 depends only weakly on the chosen masses, e.g.~the results for u and c 
 quarks differ only slightly.
 GALUGA was also used to calculate \LTT.
 \par
 Radiative corrections are calculated with the program BDK~\cite{bdk}. 
 As for the GALUGA Monte Carlo, the BDK program calculates 
 the QPM cross-section with,  in addition, initial and final state 
 QED radiative corrections to the scattered electrons.
 It has been verified that the non-radiative cross-sections predicted
 by BDK and GALUGA agree with each other. GALUGA has more flexibility  
 for calculating cross-sections and is 
 therefore used to calculate the $\LTT$ factors and QPM predictions.
 The size of the radiative corrections depends on the variables used
 to calculate the kinematics, and also to some extent on
 the non-radiative cross-section.
 \par
\begin{figure}[t]
\begin{center}
{\includegraphics[width=1.0\linewidth]{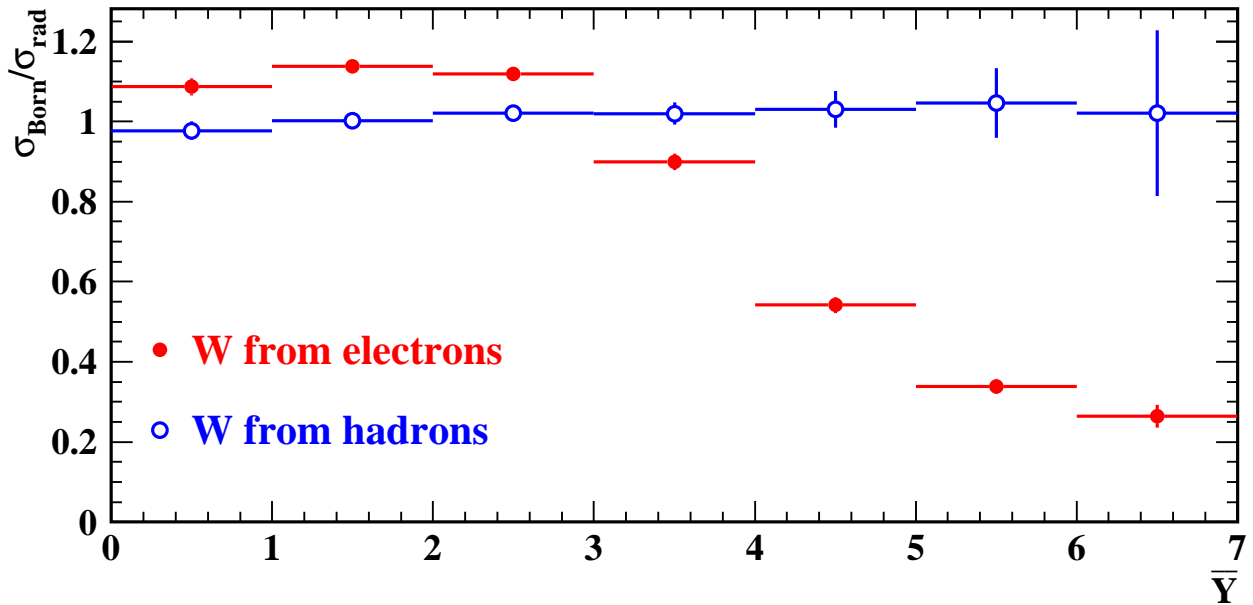}}
\caption{Radiative corrections for the process
         $\epem \rightarrow \epem \,\mbox{hadrons}$
         as a function of \Ybar for two different methods to calculate 
         $W$: a combined electron/hadronic final state method (electrons), 
         and a purely hadronic final state method (hadrons), as explained 
         in the text. Only statistical errors are shown.}
\label{radcor}
\end{center}
\end{figure}

 Fig.~\ref{radcor} compares two methods of calculating the variable \Ybar.
 The first method uses the hadronic final state to calculate $W$.
 The second method is a hybrid that calculates $W$ using $y_1$ and $y_2$ 
 obtained from the electrons, if both are positive and at least 
 one $y_i$ is larger than 0.25 otherwise \Ybar is calculated as for the 
 first method.
 The non-radiative QPM cross-section was reweighted to the leading order
 PHOJET cross-section, which agrees well with the measured cross-section 
 (see Section 7).
 Fig.~\ref{radcor} shows the ratio of the non-radiative (Born) to the full 
 radiative, detector level measured (rad) cross-section. 
 For the fully hadronic method the radiative
 corrections are small. However, for the electron method  the corrections
 can be  larger than  50\% at large \Ybar values. 
 Obviously, measurements based on the electron kinematics cannot be compared
 with models or BFKL calculations in the region $\Ybar>4$,
 unless radiative corrections have been applied.
 Since the actual size of the radiative corrections also depends on
 the non-radiative cross-section itself, an iterative procedure would 
 be required to extract the non-radiative cross-section.
 The present statistics do not permit such a procedure, so in the 
 analysis which follows \Ybar has been calculated only with the hadronic 
 variables, for which the corrections are much smaller.
%
%
\section{The OPAL detector}
\label{sec:detec}
 A detailed description of the OPAL detector can be found in 
 Ref.~\cite{opal_det}. Here only a brief account of the main components
 relevant to the present analysis is given.

The central tracking system is located inside a solenoid magnet
which provides a uniform axial magnetic field of 0.435 T along the
beam axis. The central
tracking system consists of a two-layer silicon micro-vertex 
detector~\cite{vert_det}, a high precision vertex drift chamber, a large
volume jet chamber and a set of $z$--chambers for accurately
measuring track coordinates along the beam direction. The transverse
momenta, \pT, of tracks are measured with a precision
of $\sigma_{\pT}/\pT = \sqrt{0.02^{2}+(0.0015\cdot\pT)^{2}}$ (\pT in GeV). 

The central detector is surrounded in the barrel region
($|\cos\theta | < 0.82$) by a lead glass electromagnetic calorimeter
(ECAL) and a hadronic sampling calorimeter (HCAL). Outside the HCAL,
the detector is surrounded by muon chambers. There are similar
layers of detectors in the end caps (0.81 $< |\cos\theta | <$ 0.98).
The barrel and end cap sections of the ECAL are both constructed from
lead glass blocks, with a depth of 24.6 radiation lengths in the
barrel region and more than 22 radiation lengths in the endcaps.

The small angle region from 47 to 140 mrad around the beam pipe on
both sides of the interaction point is covered by the forward
detectors (FD) and the region from 25 to 59 mrad by the silicon-tungsten 
luminometers (SW)~\cite{sw_det}. 
The lower boundary of the SW acceptance is effectively 33 mrad
due to the installation of a low-angle shield to protect the
central detector from synchrotron radiation.

The FD consists of cylindrical lead-scintillator calorimeters with a 
depth of 24 radiation lengths divided azimuthally into 16 segments.
The electromagnetic energy resolution is approximately 18\%/$\sqrt{E}$,
where $E$ is in GeV.

The SW detector consists of two cylindrical small angle calorimeters
encircling the beam pipe at approximately $\pm$2.5 m from the
interaction point. Each calorimeter is made of a stack of 18
tungsten plates, interleaved with 19 layers of silicon sampling
wafers and mounted as two interlocking C--shaped modules around
the LEP beam pipe. The depth of the detector amounts to 22
radiation lengths. Each silicon layer consists of 16
wedge-shaped silicon detectors. The sensitive area of the
calorimeter fully covers radii between 81 and 142 mm from the beam
axis. The energy resolution is approximately $5\%$
on both sides. It is found that the energy resolution is almost constant
with the energy due to energy leakage and dead material. 
In this analysis, the SW detector was used 
for tagging the scattered electrons from the process
$\epem \rightarrow \epem\,\mbox{hadrons}$.
%
%
\section{Event selection}
\label{sec:selec}
The data sample used in this analysis corresponds to an integrated
luminosity of 592.9~pb$^{-1}$ accumulated by the OPAL experiment
in 1998 (168.5~pb$^{-1}$),  1999 (208.3~pb$^{-1}$)
and 2000 (216.1~pb$^{-1}$)
at  e$^{+}$e$^{-}$ centre-of-mass energies
$\ssee=189-209$~\gev with a luminosity weighted average 
of $\sqrt{\see}$ = 198 GeV.
Double-tagged two-photon events were selected with the following set
of cuts:
\begin{enumerate}
\item Two electron candidates should be observed, one in each SW detector,
with energies $E_{1,2}>0.4\eb$  and
polar angles in the range 34 $<\theta_{1,2} <$ 55 mrad. The
angles $\theta_{1,2}$ are measured with respect to the original beam 
direction.
On each side the calorimeter cluster with the highest 
energy is taken as the electron candidate. The energy threshold 
for the electron candidates is kept as low as possible in order to 
access large $W$ values, where  BFKL effects may become significant.
\item In order to remove events with scattered electrons in FD or in
the central electromagnetic calorimeter, we require that there is no 
single cluster in these detectors with an energy above $0.25\eb$.
\item At least 3 tracks ($N_{\rm ch}\geq 3$) have to be found in the tracking system.
A track is required to have a minimum transverse momentum of 120
MeV, and to fulfill standard quality cuts as given in~\cite{opal96}.
\item The visible invariant mass, $W_{\rm vis}$, is required
to be larger than $5\,$GeV. It is reconstructed from tracks
measured in the central tracking detectors and 
the position and energy of clusters measured in
the electromagnetic and hadronic calorimeters, as well as  
in the forward detectors FD. 
A matching algorithm~\cite{opal96} is used to avoid double 
counting of the particle momenta in the calorimeters and tracking 
chambers.
\item 
 To reduce the background due to beam--gas interactions,
 the $z$ position of the primary vertex $|\langle z_{0} \rangle|$ is 
 required to be less than 4 cm from the nominal interaction point.
 Here $\langle z_{0} \rangle$ is calculated as the error weighted average of
 the $z$ coordinates of all tracks at the point of closest approach
 to the origin in the $r,\phi$ plane. 
 The standard requirement that the distance of any track to the origin of the 
 $z$ axis should be less than 30 cm was not applied in order to 
 keep all possible tracks including those from off-momentum electrons 
 interacting with the beam pipe. 
 We also require that the distance of the primary vertex from the beam axis
 should be less than 0.5 cm.
\item 
 In order to ensure that the event is well contained in the
 detector and to reduce background from beam--gas
 interactions, it is required that the $z$ component of the 
 total momentum vector of the event, $|\sum p_{\rm z}|$ is less than 35 GeV, 
 and that the total energy measured in the event is less than $2.2 \eb$.
\item
 Remaining Bhabha-like events
 (i.e. a Bhabha event with random overlap of hadronic activity) 
 are tagged using the back-to-back topology of the scattered electrons, 
 if both have an energy larger than $0.7\eb$.
 Events are rejected if the difference in radius, $\Delta r$ and 
 difference in azimuthal angle, $\Delta \phi$ of the position of the two 
 clusters in the laboratory reference frame are 
 $\vert\Delta r\vert<0.5$~cm and ($\pi -0.1)<\vert\Delta\phi\vert<\pi$~rad.
\end{enumerate}
 With these cuts 175 events are selected in the data.
 Among these we expect background events from two main sources.
 The first contribution is from \epem processes containing electrons in 
 the final state and the second stems from coincidences of \epem processes
 without electrons  in the final state with off-momentum electrons from
 beam--gas interactions.
 \par
 The background from \epem reactions containing electrons in the final
 state, which amounts to 18.3 events, dominantly stems from the processes 
 $\epem\rightarrow\epem\tau\overline{\tau}$ and $\epem \rightarrow \epem\epem$,
 and was estimated using the Vermaseren Monte Carlo program~\cite{vermas}.
 The contribution from other $\epem$ background, such as $\epem 
 \rightarrow q\overline{q}$ and other  processes leading to four
 fermion final states was found to be negligible, as was the background from 
 single-tagged two-photon processes with a cluster from the 
 hadronic final state misidentified as the second electron.
 \par
 Beam--gas interactions can result in off-momentum electrons observed
 in the SW detectors, faking final state electrons from the process
 $\epem \rightarrow \epem\,\mbox{hadrons}$.
 This background was estimated using a sample of Bhabha events, 
 selected by requiring events with two back-to-back electrons in the SW 
 calorimeters, which each have an energy of more than 0.7$\eb$, with 
 $\vert\Delta r\vert < 0.5$~cm and ($\pi-0.1)<\vert\Delta\phi\vert<\pi$~rad. 
 Additional clusters in the SW detectors, which fulfill the criteria for 
 electron candidates mentioned above, but do not belong to the Bhabha event,
 are counted as off-momentum electrons.
 The probabilities to have an overlapping off-momentum electron with an 
 event coming from the interaction region are determined for the 
 left ($-z$) / right ($+z$) side of the detector separately and amount to 
 0.000715/0.00115 (1998), 
 0.00139 /0.00279 (1999) and 
 0.00117 /0.000671 (2000).
 The relative statistical precision of these probabilities is 2-3\%.
 The background expectation is more than a factor two larger in 1999 than
 in the other years.
 Using these numbers, and assuming that the overlap probabilities are 
 independent between the left and right side of the detector, we 
 predict that there should be 14.6 events with 'double' overlaps in the 
 Bhabha sample, which agrees well with the 19 events observed.
 \par
 It is essential to check this method of estimating the background from 
 off-momentum electrons on a different process. Here we used a sample of
 single-tagged two-photon events.
 The sample is selected with the same cuts as described for the 
 double-tagged selection, except that only one scattered electron is required
 and cuts 6) and 7) are not applied.
 The event properties are compared with the absolute prediction for the 
 single-tagged plus background events.
 The single-tagged events are based on the HERWIG generator with the 
 GRV~\cite{grv} parametrisation of the photon structure function $F_2$,
 which has been shown to describe the single-tagged 
 cross-section~\cite{OPALPR314} within about 10$\%$ for $E > 0.7\eb$. 
 The genuine single-tagged events are complemented by the physics background 
 and by artificially created 
 single-tagged events constructed by a combination of anti-tagged two-photon
 events generated by PHOJET with a cluster created by an off-momentum
 electron.
 The $\theta$, $\phi$ and energy dependence of the electron clusters
 are given by the spectra of the additional clusters in Bhabha events.
 
Fig.~\ref{overlap1} shows the 1999 data, which is expected to 
have the largest background from off-momentum overlaps, compared with
the prediction
resulting from the sum of HERWIG and the background, normalised
to the luminosity of the data.
 Results are shown for $\theta$, $\phi$ and energy of the scattered 
 electron normalised to the energy of the beam electrons.
 Fig.~\ref{overlap3} shows separately for the 1998, 1999 and 2000 data samples,
 the missing longitudinal momentum and missing transverse momentum in the
 event, calculated including the untagged electron,
 which has been assumed to have zero transverse momentum and an energy
 equal to the energy of the beam electrons.  
 Both distributions have been normalised to the energy of the beam electrons. 
 For energies below $0.6\eb$ the off-momentum background clearly dominates
 and the observed angular dependences, especially in $\phi$, clearly
 follow the expected shape.

The agreement between data and prediction is very good for all variables
examined, providing confidence that the background from overlap off-momentum
beam electrons is under control to a level of about 10\%.
The off-momentum background estimate was used to calculate the contribution 
of fake double-tagged events, resulting from the overlap of one background
cluster with a single-tagged two-photon event and the overlap 
of two background clusters with an untagged event. 
 In total 4.3, 15.2 and 4.6 overlap events are predicted
 for the 1998, 1999 and 2000 data samples.
 After subtraction of all backgrounds 132.6 events remain, with each year
 contributing with 31.9$\pm$6.5 (1998), 63.5$\pm$9.3 (1999) and
 37.2$\pm$7.0 (2000) events, whereas 42.5, 53.7 and 55.5 events are predicted
 by PHOJET. 
 The difference between the number of events observed in the three years 
 is of the order of twice their statistical error.
 Since the statistics of this measurement are small,
 the data of all years will be combined in the following.
 \par
 The double-tagged events are triggered by two groups of independent triggers.
 The first trigger is based on the energy deposits of the observed electrons in
 the SW calorimeters.
 The second group only relies on the observed tracks and clusters from the
 hadronic final state.
 Based on these two independent groups, the trigger efficiency of the hadronic
 final state alone, for events with two electrons with energies above 0.4$\eb$,
 has been determined from the data to be $(96\pm4)\%$.
%
%
\section{Properties of double-tagged \boldmath\ggss\unboldmath events}
\label{sec:prop}
A PHOJET Monte Carlo sample  is used to correct the data for acceptance
and resolution effects. It is therefore essential that the shape of all
important distributions is well reproduced by the Monte Carlo simulation. 
In this Section a comparison is made of data distributions with predictions
from PHOJET. Variables calculated from the scattered electrons as well as
variables calculated from the hadronic final state are studied.
The integrated luminosity of the Monte Carlo sample amounts to approximately
$40$ times that of the data. All Monte Carlo distributions shown in this
Section are normalised to the data luminosity. In all plots involving both
sides of the SW detector,  the sum of the distributions obtained separately
for each side is shown.

In Fig.~\ref{comp_ele} variables which are based on electron quantities,
and the variables $x_{1,2}$, are compared 
with predictions of PHOJET and background estimates.
All variables, the normalised electron energies $E_{1,2}$, the polar angles
$\theta_{1,2}$, the azimuthal angles $\phi_{1,2}$, and photon
virtualities $Q^2_{1,2}$, are reasonably well described by the sum of the 
signal as predicted by PHOJET and the estimated background from overlaps with 
off-momentum electrons and other physics processes. 
 Note that PHOJET does not contain any explicit effects from BFKL, which 
 would show up in the region of low electron energies. 
 Fig.~\ref{comp_ele}e) shows the logarithm of the ratio of the photon
 virtualities, $\ln (Q^2_1/Q^2_2)$, of the two photons in an event.
 This distribution is peaked around zero, indicating that the 
 $Q^2$ values of both photons are generally close to one another, which is 
 ideal to test for BFKL effects. 
 \par
 In Fig.~\ref{comp_had} distributions are shown which characterise the hadronic
 final state in double-tagged two-photon events: the
 number of tracks, $N_{\rm ch}$, the visible hadronic invariant
 mass, $W_{\rm vis}$, the hadronic energy, $E_{\rm had}$, the variable \Ybar
 as well as the sum of the longitudinal $\sum p_{\rm z}$ and 
 transverse $\sum \pT$ momenta.
 Within  statistics, the agreement with PHOJET is reasonable.
%
%
\section{Results}
 The cross-section for the process $\epem \rightarrow \epem\,\mbox{hadrons}$
 has been measured in the kinematic region defined by the scattered
 electron energies $E_{1,2}>0.4\eb$, the polar angles in the range
 $34<\theta_{1,2}<55$ mrad with respect to either beam
 direction, and $W>5$ GeV.
 Differential cross-sections are presented as a function of $x$,
 $Q^2$, $W$, and the azimuthal correlation between the two electrons
 $\Delta\phi$. 
 Here $Q^{2}$ refers to the maximum of $Q^{2}_{1}$ and $Q^{2}_{2}$, 
 and $x$ is the corresponding value of $x_1$ or $x_2$.
 For the comparison with BFKL predictions we also present
 the differential cross-section as a function of \Ybar.
 From the measurement of the differential cross-sections of the process
 $\epem \rightarrow  \epem\,\mbox{hadrons}$ we extract the cross-sections 
 for the reaction $\ggss \rightarrow \mbox{hadrons}$ as a function of the
 variable under study, using \LTT, obtained 
 separately for each bin using Monte Carlo.
 Technically this was done by setting $\sgg\equiv 1$ in GALUGA 
 and integrating (Eq.~\ref{eqn:truesimp}) using only \LTT
 for each bin within the experimental phase space restrictions.
 The cross-sections for the reaction $\ggss \rightarrow\,\mbox{hadrons}$
 predicted by the models are calculated using the same $\LTT$ factors.
 For all variables except \qsq the results for the cross-sections of 
 the reaction $\ggss \rightarrow\,\mbox{hadrons}$
 are given at an average value of \qsq, $\langle Q^2 \rangle = 17.9$~\gevsq.
 Due to limited statistics in the data, a simple bin-by-bin
 method was applied to correct for detector and
 selection inefficiencies. The efficiency, $R_{\rm e}$, and purity,
 $R_{\rm p}$, are defined as:
 
 $$R_{\rm e}=\frac{N^{\rm Det \otimes Had}}{N^{\rm Had}} \;\;\;\;\;\;\;\;\;
 R_{\rm p}=\frac{N^{\rm Det \otimes Had}}{N^{\rm Det}} $$
where $N^{\rm Det \otimes \rm Had}$ is the number of events which are generated
in a bin and measured in the same bin, $N^{\rm Had}$ is the number of events 
which are generated in a bin and $N^{\rm Det}$ is the number of events
measured in a bin. In both definitions the terms `generated' and `measured'
denote events which pass all selection cuts at the hadron or at the detector
level, respectively. 
The correction factor $N^{\rm Had}/N^{\rm Det}$ is 
obtained by dividing purity by efficiency.
For the $W$ variable the purity is typically around 60\% over the whole
range, and the efficiency is in the range of 30-50\%.
Similar numbers are obtained for \Ybar and $x$, while for the  $\Delta\phi$
and $Q^2$ variables the efficiencies are around 60\% and purities around 80\%.
The correction factor is typically around 1.5 and fairly constant.

 The systematic error of the measurement has been evaluated taking into
 account several contributions. 
 All changes, except the change of the cut on $N_{\rm ch}$, are applied to
 the Monte Carlo event samples because of low statistics of the data.
\begin{enumerate}
\item The error due to a possible shift of the energy scale of the SW detectors
      was taken into account by scaling the electron energy by $\pm$1\%,
      in accord with the uncertainty in the scale observed in single-tagged 
      events, which is conservative because the differences observed
      for Bhabha events were much smaller.
\item The uncertainty in the description of the energy scale of the central
      electromagnetic calorimeter ECAL was taken into account by varying the
      energy scale by $\pm 3\%$~\cite{OPALPR314}.
\item To estimate the uncertainties due to the selection cuts, we
      have performed the following checks, reflecting the resolution
      of the variables:
\begin{enumerate}
\item The lower cut on $\theta_{1,2}$ was changed from 34 mrad 
      by $\pm0.4$ mrad.
\item The lower cut on $E_{1,2}$ was changed by $\pm 5\%$.
\item The cut on $W$ was changed by $\pm 0.5$ GeV.
\item The cut on $|\langle z_{0} \rangle|$ was changed by $\pm  1.0$ cm.
\item The cut on the distance of the primary vertex from the beam 
      axis was changed by $\pm 0.1$ cm.
\end{enumerate}
\item The PYTHIA Monte Carlo was used instead of PHOJET to correct the data.
      The differences between the models are generally within 10-15$\%$, 
      except for the high $W$ region where the difference amounts to 25$\%$.
      The full differences were taken as the errors.
\item In modelling the region of $W<5$ GeV PHOJET is incomplete. This 
      affects the measurement only through migrations from lower to
      higher values of $W$. To estimate the influence of the 
      imperfect modelling of that region in 
      PHOJET we have chosen the following method.
      The cross-section for events with $W<5$~GeV have been rescaled such that 
      there was a smooth transition in the $W$ distribution around $W=5$~GeV. 
      Then the number of events generated for $W<5$~GeV were varied by 
      $\pm 50\%$.
\item The uncertainty in the estimation of the off-momentum background was 
      taken to be 10\% of the background, motivated by the level of agreement
      observed for the single-tagged events.
\item The cut on the number of charged tracks has been changed from
      $N_{\rm ch}\geq 3$ to $N_{\rm ch}\geq 4$. This change has been applied
      simultaneously to data and Monte Carlo.
\item The 4\% uncertainty in the trigger efficiency was added to the total
      systematic error.
\end{enumerate}
 The total systematic error was obtained by adding in quadrature
 errors coming from the above checks, where the sum has been performed
 separately for positive and negative contributions.
 The main contributions to the systematic errors come from changing the
 Monte Carlo model from PHOJET to PYTHIA 
 ($-16\%$ change in the total cross-section),
 changing the cut on $N_{\rm ch}$ ($-12\%$), and varying the lower cuts on
 $W$ ($\pm 8\%$) and $\theta_{1,2}$ ($\pm 5\%$).
 The  normalisation uncertainty due to the luminosity measurement is less
 than 1$\%$ and has been neglected.
 \par
 The total measured cross-section for the process 
 $\epem \rightarrow \epem\,\mbox{hadrons}$ in the previously defined phase 
 space, is $0.35 \pm 0.04$ (stat) $^{+0.04}_{-0.08}$ (sys) pb. 
 The expected cross-section from PHOJET is $0.39\pm0.02({\rm stat})$~pb, 
 while the GALUGA prediction for QPM using massive quarks is 
 $0.27\pm0.02 ({\rm stat})$~pb, and the NLO predictions for the reaction
 \eeqq using massless quarks is 0.35~pb.
 In Fig.~\ref{cros11} we show the measured differential cross-section for
 the processes $\epem \rightarrow  \epem\,\mbox{hadrons}$ and the 
 cross-section for $\ggss \rightarrow\,\mbox{hadrons}$ as a function of $x$
 and \qsq.
 The numerical values are given in Tables~\ref{t_db} and~\ref{t_dbx}.
 PHOJET better describes the data at lower $x$ values, where the QPM 
 prediction is too low. In contrast, in the large $x$ region 
 the QPM prediction is sufficient to account for the data.
 \par
 Fig.~\ref{cros12} shows the measured cross-sections as a 
 function of $W$ and  $\Delta \phi$.
 The  model  predictions indicate a slightly different shape than 
 is observed for the data.
 Studies for HERA have shown~\cite{delducca} that
 angular variables similar to $\Delta \phi$ can be sensitive
 to the presence of BFKL dynamics, but so far  no calculations are
 available for \ggss scattering.
 The data show that the $\Delta \phi$ behaviour of the cross-section
 for the reaction $\epem\rightarrow\epem\,\mbox{hadrons}$
 is flat, while the cross-section for the process
 $\ggss\rightarrow\,\mbox{hadrons}$ increases from 
 $\Delta\phi=0$ to $\Delta\phi=\pi$.
 PHOJET1.10 does not describe the $\Delta \phi$ distribution, whereas
 QPM reproduces the shape of the distribution.
 One should remark that the earlier version, PHOJET1.05, follows
 the data in both shape and normalisation.
 \par
 In Fig.~\ref{cros13} we compare the measured cross-section 
 for the processes $\epem \rightarrow  \epem \,\mbox{hadrons}$ and
 $\ggss \rightarrow\,\mbox{hadrons}$ as a function of \Ybar
 with the PHOJET Monte Carlo, the QPM calculation, the
 NLO calculation for the reaction \eeqq, and 
 numerical BFKL calculations (Tables~\ref{t_db} and~\ref{t_dbx}).
 The BFKL predictions are shown for the LO-BFKL~\cite{carlo}
 and a NLO-BFKL~\cite{kim} calculation both using $Y$. 
 Also shown is a (partial) HO-BFKL calculation~\cite{kwiecinski}
 using \Ybar.
 All BFKL predictions are shown for $Y\,{\mathrm or}\,\Ybar>2$, except 
 for the NLO-BFKL calculation which has been evaluated for $Y>1$.
 Both PHOJET and QPM using massive quarks describe the data equally well.
 Also the NLO calculation for the reaction \eeqq, evaluated for 
 five massless quarks and using $\lamf=0.2275$~\gev, is in accord with the 
 data.
 As can be seen from Table~\ref{t_nlo}
 the predicted differential cross-sections as functions of $Y$ and \Ybar
 are very different at small values, but get much closer at higher values.
 This is expected from the approximation made in Eq.~\ref{eqn:ybar}.
 It means that at low values of \Ybar the comparison of the experimental
 result with predictions based on $Y$ is rather uncertain, whereas as high
 values the uncertainty from using different definitions is small.
 \par
 For all BFKL predictions shown the cross-section is significantly
 larger than the PHOJET prediction for $\Ybar>3$, and the differences
 increase with increasing $\Ybar$. The LO-BFKL calculation predicts a
 cross-section which is too large compared to the data. This LO-BFKL
 calculation (Bartels99)~\cite{carlo} already incorporates improvements
 compared to the original results~\cite{bartels} by including effects
 of the charm quark mass, the running of the strong coupling constant
 $\alpha_s$ and contribution of longitudinal photon polarisation states.
 Hence BFKL effects as large as predicted by the LO-BFKL calculation 
 are not in agreement with the data.
 BFKL cross-sections have been calculated to NLO (Kim99)~\cite{kim},
 using the BLM~\cite{blm} optimal scale setting.
 At the highest \Ybar value the NLO-BFKL cross-section is a factor
 seven larger than the PHOJET prediction. The data lie in between
 these two predictions. 
 Finally, the calculation (Kwiecinski)~\cite{kwiecinski} contains
 the dominant contribution of the higher order corrections via the
 so-called consistency constraint, to all orders.
 Its prediction in the highest reachable \Ybar
 range is about a factor two lower than for the NLO-BFKL calculation, and
 this prediction and PHOJET are roughly equally compatible with the data
%
%
\section{Summary and conclusions}
\label{sec:concl}
 A data sample collected by the OPAL experiment at LEP based on an
 integrated luminosity of 592.9~pb$^{-1}$ and for \epem centre-of-mass
 energies $\ssee=189-209$~\gev has been used to study interactions of
 virtual photons.
 Differential cross-sections for the processes 
 $\epem\rightarrow\epem\,\mbox{hadrons}$ are measured as functions of
 $x$, \qsq, $W$, $\Delta\phi$ and \Ybar.
 From this, the cross-sections for the reaction 
 $\ggss \rightarrow\,\mbox{hadrons}$ as a functions of these variables
 have been derived.
 Both PHOJET1.10 and QPM describe the data equally well for
 the cross-sections in $x$, \qsq, $W$, and \Ybar.
 PHOJET1.10 does not describe the $\Delta \phi$ distribution, whereas
 QPM reproduces the shape of the distribution.
 Also the NLO calculation for the reaction \eeqq is in accord with the data.
 \par
 Within current calculations BFKL effects could lead to an increase of
 the cross-section by a factor up to 20 for the largest \Ybar values.
 The data rule out BFKL cross-sections which are as large as those
 predicted by LO-BFKL and NLO-BFKL calculations as presented
 in~\cite{kim}. Calculations including dominant higher order
 corrections predict smaller effects in the LEP range and are found to
 be consistent with the measured cross-sections.
 The limited statistics and available \Ybar range of the 
 data prevent establishing or ruling out the onset of BFKL dynamics 
 in this reaction.
%
%
 \clearpage
 \appendix
 \par 
\section*{Acknowledgements:}
 \par
  We are grateful to C.~Ewerz, V.~Kim, J.~Kwiecinski and L.~Motyka for
  providing their predictions. 
  We wish to thank S.~Frixione for providing the software to calculate the 
  NLO predictions.
 \par
 We particularly wish to thank the SL Division for the efficient operation
 of the LEP accelerator at all energies
 and for their continuing close cooperation with
 our experimental group.  We thank our colleagues from CEA, DAPNIA/SPP,
 CE-Saclay for their efforts over the years on the time-of-flight and
 trigger systems which we continue to use.  In addition to the support 
 staff at our own institutions we are pleased to acknowledge the  \\
 Department of Energy, USA, \\
 National Science Foundation, USA, \\
 Particle Physics and Astronomy Research Council, UK, \\
 Natural Sciences and Engineering Research Council, Canada, \\
 Israel Science Foundation, administered by the Israel
 Academy of Science and Humanities, \\
 Minerva Gesellschaft, \\
 Benoziyo Center for High Energy Physics,\\
 Japanese Ministry of Education, Science and Culture (the
 Monbusho) and a grant under the Monbusho International
 Science Research Program,\\
 Japanese Society for the Promotion of Science (JSPS),\\
 German Israeli Bi-national Science Foundation (GIF), \\
 Bundesministerium f\"ur Bildung und Forschung, Germany, \\
 National Research Council of Canada, \\
 Research Corporation, USA,\\
 Hungarian Foundation for Scientific Research, OTKA T-029328, 
 T023793 and OTKA F-023259.\\
%
%

%
\clearpage
%
\renewcommand{\arraystretch}{1.15}
\begin{table}[ht]
 \begin{center}\begin{tabular}{|c|c|c|c|c|}\hline
 range & \multicolumn{2}{|c|}{${\der\sigma} / {\der \Ybar}$} 
       & \multicolumn{2}{|c|}{${\der\sigma} / {\der Y}$} \\
\cline{2-5}
       &          LO & NLO           &       LO & NLO            \\
\hline
 0 -- 1 & 0.071 & 0.065 & 0.015 & 0.014 \\
 1 -- 2 & 0.135 & 0.128 & 0.149 & 0.138 \\
 2 -- 3 & 0.087 & 0.089 & 0.111 & 0.122 \\
 3 -- 4 & 0.041 & 0.047 & 0.047 & 0.053 \\
 4 -- 6 & 0.010 & 0.013 & 0.011 & 0.014 \\
\hline
\end{tabular}
\caption{Predictions for the differential cross-section for the process \eeqq
         as functions of \Ybar and $Y$ using the calculation
         from~\protect\cite{Cacciari}.
        }\label{t_nlo}
\end{center}\end{table}
%
%
\renewcommand{\arraystretch}{1.15}
\begin{table}[ht]
\begin{tabular}{|c|c|c|c|c|c|c|c|}
\hline
$\langle x\rangle$ & range & $N_{\rm ev}$ &
$\frac{\der\sigma}{\der x}$ OPAL &
Statistical & \multicolumn{2}{|c|} {Systematic error} &
$\frac{\der\sigma}{\der x}$ PHOJET \\
\cline{6-7}  &  &  & [pb] & error & ~~~up~~~ & down & [pb] \\
\hline
0.06 & 0.0 - 0.1~ & 45.6 & 1.46 & 0.27 & 0.11 & 0.26 & 1.29 \\
0.15 & 0.1 - 0.2~ & 50.1 & 1.22 & 0.20 & 0.10 & 0.19 & 1.26 \\
0.26 & 0.2 - 0.35 & 34.9 & 0.55 & 0.11 & 0.11 & 0.20 & 0.85 \\
\hline\multicolumn{8}{c}{}\\\hline
$\langle Q^{2}\rangle$ & range & $N_{\rm ev}$ & $\frac{\der\sigma}{\der\qsq}$
OPAL & Statistical & \multicolumn{2}{|c|} {Systematic error} &
$\frac{\der\sigma}{\der\qsq}$ PHOJET \\
\cline{6-7} [GeV$^{2}$] & [GeV$^{2}$] &  & [pb/GeV$^{2}$] & error &
~~~up~~~ & down & [pb/GeV$^{2}$] \\
\hline
13.6 & 10 - 16 & 32.3 & 0.016 & 0.003 & 0.002 & 0.003 & 0.018 \\
18.9 & 16 - 22 & 59.4 & 0.026 & 0.004 & 0.003 & 0.006 & 0.027 \\
24.4 & 22 - 27 & 25.0 & 0.013 & 0.003 & 0.001 & 0.003 & 0.020 \\
\hline\multicolumn{8}{c}{}\\\hline
$\langle W \rangle$ & range & $N_{\rm ev}$ & $\frac{\der\sigma}{\der W}$ OPAL &
Statistical & \multicolumn{2}{|c|} {Systematic error} &
$\frac{\der\sigma}{\der W}$ PHOJET \\
\cline{6-7} [GeV] & [GeV] &  & [pb/GeV] & error & ~~~up~~~ & down &
[pb/GeV] \\
\hline
~7.2 & ~5 - 10 & 65.1 & 0.031 & 0.004 & 0.0053 & 0.0098 & 0.041 \\
12.4 & 10 - 15 & 30.6 & 0.017 & 0.004 & 0.0012 & 0.0015 & 0.021 \\
20.6 & 15 - 35 & 32.5 & 0.005 & 0.001 & 0.0004 & 0.0010 & 0.004 \\
41.5 & 35 - 50 & ~3.4 & 0.001 & 0.001 & 0.0001 & 0.0001 & 0.000 \\
\hline\multicolumn{8}{c}{}\\\hline
$\Delta\phi$ & range & $N_{\rm ev}$ & ~$\frac{\der\sigma}{\der\Delta\phi}$
OPAL &
Statistical & \multicolumn{2}{|c|} {Systematic error} &
$\frac{\der\sigma}{\der\Delta\phi}$ PHOJET \\
\cline{6-7} [rad] & [rad] &  & [pb/rad] & error & ~~~up~~~ & down & [pb/rad] \\
\hline
0.31 & $ 0.00$ - $ 0.63$ & 20.2 & 0.081 & 0.022 & 0.010 & 0.032 & 0.195 \\
0.94 & $ 0.63$ - $ 1.26$ & 23.7 & 0.098 & 0.024 & 0.012 & 0.035 & 0.163 \\
1.57 & $ 1.26$ - $ 1.89$ & 31.2 & 0.130 & 0.026 & 0.013 & 0.036 & 0.120 \\
2.20 & $ 1.89$ - $ 2.51$ & 27.1 & 0.117 & 0.026 & 0.011 & 0.020 & 0.083 \\
2.83 & $ 2.51$ - $ 3.14$ & 30.4 & 0.132 & 0.028 & 0.009 & 0.010 & 0.068 \\
\hline\multicolumn{8}{c}{}\\\hline
$\Ybar$ & range & $N_{\rm ev}$ & ~$\frac{\der\sigma}{\der \Ybar}$ OPAL &
Statistical & \multicolumn{2}{|c|} {Systematic error} &
$\frac{\der\sigma}{\der \Ybar}$ PHOJET \\
\cline{6-7}  &  &  & [pb] & error & ~~~up~~~ & down & [pb] \\
\hline
0.5 & 0 - 1 & 25.7 & 0.059 & 0.013 & 0.029 & 0.033 & 0.083 \\
1.5 & 1 - 2 & 44.9 & 0.105 & 0.018 & 0.007 & 0.021 & 0.141 \\
2.5 & 2 - 3 & 37.1 & 0.108 & 0.020 & 0.009 & 0.011 & 0.111 \\
3.5 & 3 - 4 & 12.7 & 0.040 & 0.014 & 0.003 & 0.009 & 0.040 \\
5.0 & 4 - 6 & 11.1 & 0.026 & 0.010 & 0.003 & 0.010 & 0.009 \\
\hline
\end{tabular}
\caption{The differential cross-section for the process 
 $\epem\rightarrow \epem\,\mbox{hadrons}$ in the region 
 $E_{1,2}>0.4\eb$, $34<\theta_{1,2}<55$ mrad and $W>5$ GeV, 
 as a function of $x$, $Q^{2}$, $W$, $\Delta \phi$ and \Ybar.
 The average value of $x$, $Q^{2}$, $W$ and the central value for 
 $\Delta\phi$ and \Ybar in a bin, value of the bin boundaries, number
 of measured events in the bin after background subtraction, value of
 the differential cross-section with statistical and systematic errors as 
 well as the differential cross-section predicted by PHOJET1.10, are given.}
\label{t_db}
\end{table}
\renewcommand{\arraystretch}{1.15}
\begin{table}[ht]
\begin{center}
\begin{tabular}{|c|c|c|c|c|c|c|c|}
\hline
$\langle x\rangle$ &
$\sigma$ OPAL &
Statistical & \multicolumn{2}{|c|} {Systematic error} &
$\sigma$ PHOJET & $\sigma$ QPM & \LTT\\
\cline{4-5} & [nb] & error & ~~~up~~~ & down & [nb] & [nb] & $10^{-3}$\\
\hline
0.06 & 4.66 & 0.85 & 0.33 & 0.83 & 4.11 & 2.79 & 0.0314\\
0.15 & 8.92 & 1.42 & 0.71 & 1.39 & 9.18 & 6.23 & 0.0137\\
0.26 & 5.76 & 1.10 & 1.18 & 2.10 & 8.95 & 6.03 & 0.0143\\
\hline\multicolumn{8}{c}{}\\\hline
$\langle Q^{2}\rangle$ & $\sigma$ OPAL & Statistical &
\multicolumn{2}{|c|} {Systematic error} &
$\sigma$ PHOJET & $\sigma$ QPM & \LTT\\
\cline{4-5} [GeV$^{2}$] & [nb] & error &
~~~up~~~ & down & [nb] & [nb] &$10^{-3}$\\
\hline
13.6 & 5.63 & 1.23 & 0.77 & 1.05 & 6.30 & 4.61 & 0.0169\\
18.9 & 6.37 & 0.91 & 0.75 & 1.56 & 6.66 & 4.41 & 0.0241\\
24.4 & 3.91 & 0.89 & 0.36 & 0.98 & 5.93 & 4.20 & 0.0165\\ 
\hline\multicolumn{8}{c}{}\\\hline
$\langle W \rangle$ & $\sigma$ OPAL &
Statistical & \multicolumn{2}{|c|} {Systematic error} &
$\sigma$ PHOJET & $\sigma$ QPM & \LTT\\
\cline{4-5} [GeV] & [nb] & error & ~~~up~~~ & down &
[nb] & [nb]  &$10^{-3}$ \\
\hline
~7.2 & 6.79 & 0.96 & 1.17 & 2.15 & 9.03 & 6.27 & 0.0227\\
12.4 & 7.11 & 1.51 & 0.52 & 0.63 & 9.08 & 5.38 & 0.0117\\
20.6 & 5.53 & 1.14 & 0.38 & 1.11 & 4.01 & 2.95 & 0.0183\\
41.5 & 3.10 & 2.06 & 0.29 & 0.43 & 1.38 & 1.27 & 0.0045\\
\hline\multicolumn{8}{c}{}\\\hline
$\Delta\phi$ & ~$\sigma$ OPAL &
Statistical & \multicolumn{2}{|c|} {Systematic error} &
$\sigma$ PHOJET &
$\sigma$ QPM &\LTT\\
\cline{4-5} [rad] & [nb] & error & ~~~up~~~ & down & [nb] & [nb] & $10^{-3}$\\
\hline
0.31 & 3.18 & 0.84 & 0.40 & 1.25 & 7.64 & 2.82 & 0.0161\\ 
0.94 & 4.25 & 1.03 & 0.51 & 1.53 & 7.05 & 3.51 & 0.0145\\
1.57 & 6.84 & 1.38 & 0.68 & 1.88 & 6.28 & 4.74 & 0.0120\\
2.20 & 7.44 & 1.67 & 0.70 & 1.25 & 5.31 & 5.94 & 0.0099\\
2.83 & 9.56 & 2.02 & 0.68 & 0.76 & 4.90 & 6.44 & 0.0087\\
\hline\multicolumn{8}{c}{}\\\hline
\Ybar & ~$\sigma$ OPAL &
Statistical & \multicolumn{2}{|c|} {Systematic error} &
$\sigma$ PHOJET & $\sigma$ QPM & \LTT\\
\cline{4-5}  & [nb] & error & ~~~up~~~ & down & [nb] & [nb] &$10^{-3}$\\
\hline
0.5 & 6.11 & 1.34 & 2.95 & 3.42 & 8.60 & 5.37 & 0.0097\\
1.5 & 6.84 & 1.17 & 0.43 & 1.39 & 9.14 & 6.59 & 0.0154\\
2.5 & 7.99 & 1.51 & 0.69 & 0.83 & 8.27 & 4.98 & 0.0135\\
3.5 & 3.78 & 1.38 & 0.32 & 0.84 & 3.79 & 2.98 & 0.0105\\
5.0 & 4.97 & 1.82 & 0.49 & 1.81 & 1.71 & 1.40 & 0.0106\\
\hline
\end{tabular}
\end{center}
\caption{The cross-section for the process 
$\ggss \rightarrow\,\mbox{hadrons}$
as a function of $x$, $Q^{2}$, $W$, $\Delta\phi$ and $\Ybar$. 
The average value of $x$, $Q^{2}$, $W$ and the central value for 
$\Delta\phi$ and $\Ybar$ in a bin, value of
the cross-section with statistical and systematic errors,
the cross-section predicted by PHOJET1.10 and QPM as well
as \LTT are given.}
\label{t_dbx}
\end{table}
%
%
\begin{figure}[t]
\begin{center}
{\includegraphics[width=0.98\linewidth]{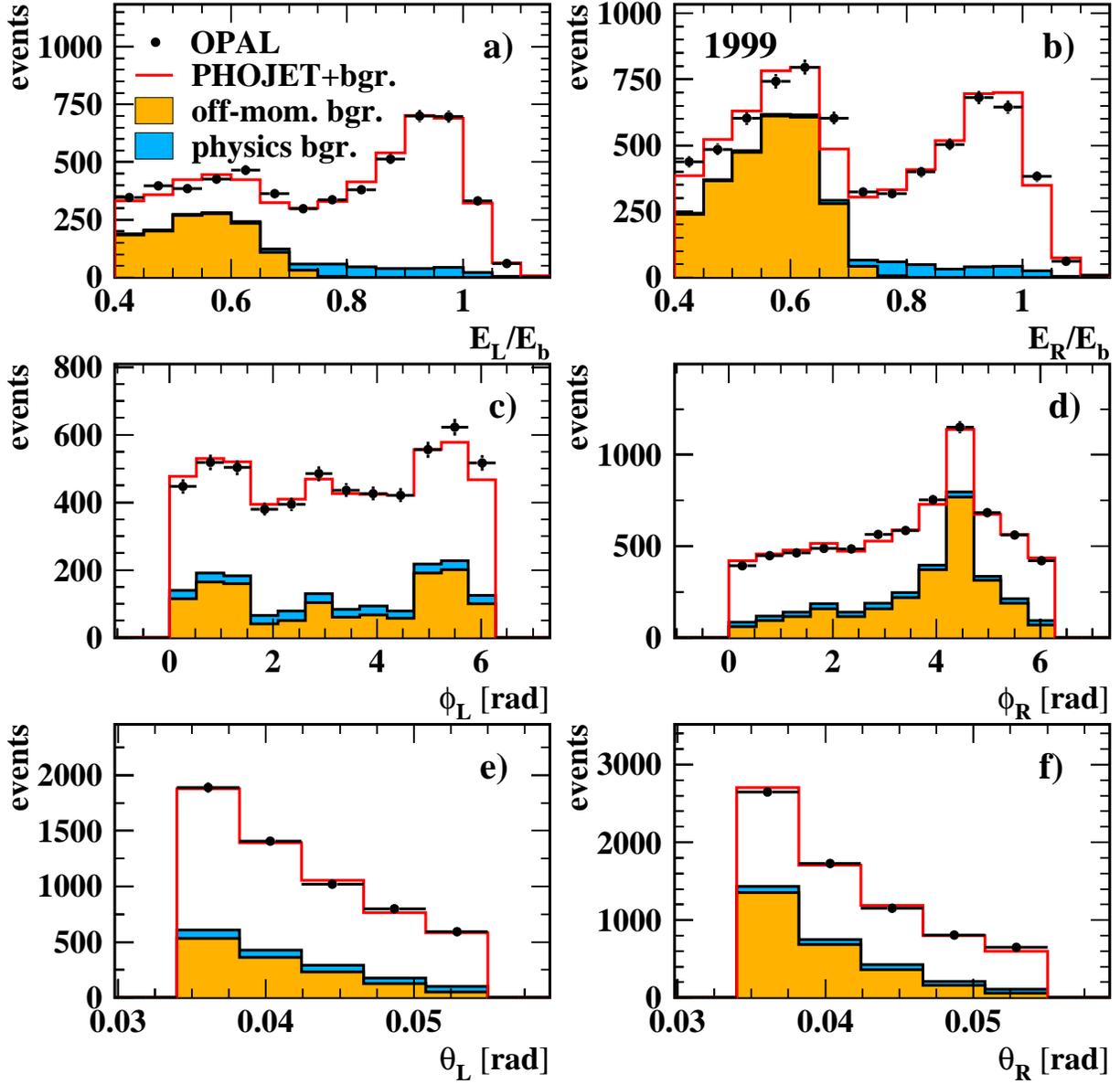}}
\caption{Distributions of the electron energy normalised to the 
energy of the beam electrons (a,b), 
electron azimuthal angle (c,d),
and electron polar angle (e,f), shown separately 
for the left (L) and right (R) side of the OPAL detector,
for selected single-tagged events in 1999. The histograms are 
the predictions for the single-tagged process from HERWIG, the 
off-momentum background contribution, and 
the background from other physics channels.}
\label{overlap1}
\end{center}
\end{figure}

\begin{figure}[t]
\begin{center}
{\includegraphics[width=0.98\linewidth]{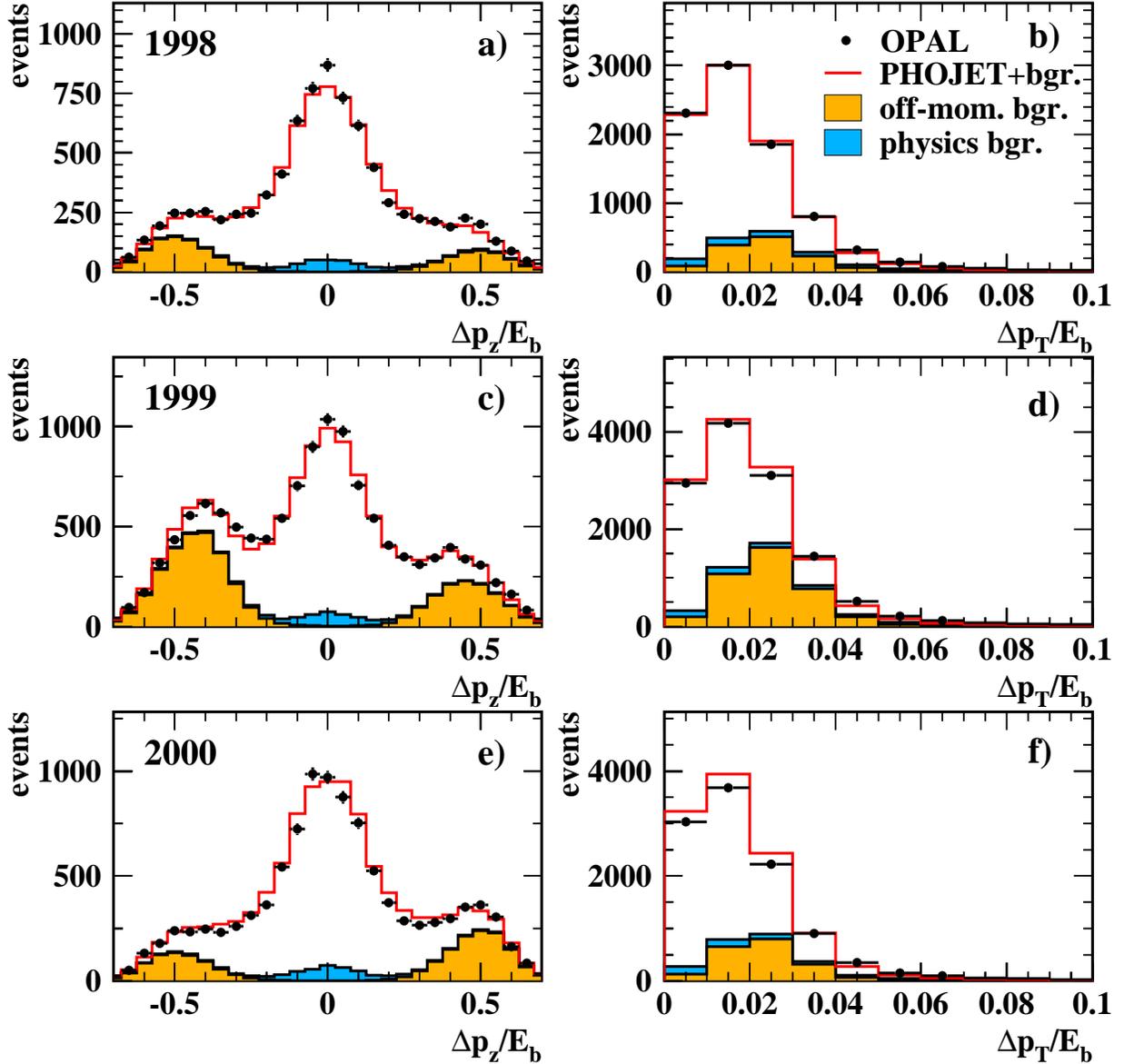}}
\caption{Distributions of the scaled missing longitudinal 
momentum (a,c,e) and the scaled missing transverse momentum (b,d,f)
in single-tagged events for the different 
years. The meaning of the histograms are as defined in Fig.~\ref{overlap1}.}
\label{overlap3}
\end{center}
\end{figure}

\begin{figure}[t]
\begin{center}
{\includegraphics[width=0.98\linewidth]{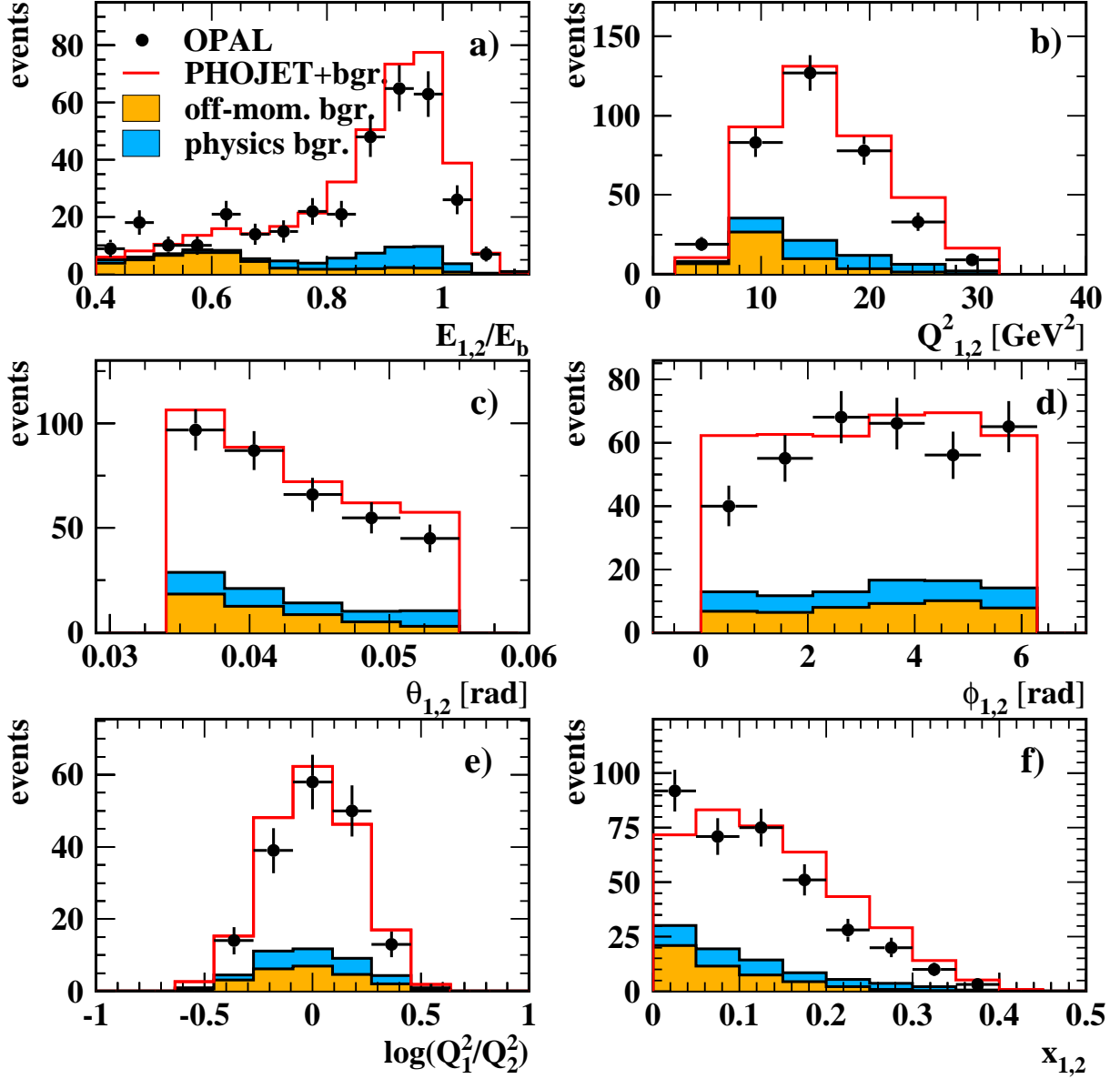}}
\caption{Distributions of  
        (a) the energies of the electrons, normalised to the 
            energy of the beam electrons, 
        (b) the  virtualities of both photons,
        (c) the polar angles of the electrons, 
        (d) the azimuthal angles of the electrons,  
        (e) the ratio of the photon virtualities and 
        (f) the $x$ values  
        of double-tagged two-photon events.
        The histograms are the predictions for the double-tagged two-photon 
        process from PHOJET1.10, the off-momentum background contribution, and 
        the background from other physics channels.}
\label{comp_ele}
\end{center}
\end{figure}

\begin{figure}[t]
\begin{center}
{\includegraphics[width=0.98\linewidth]{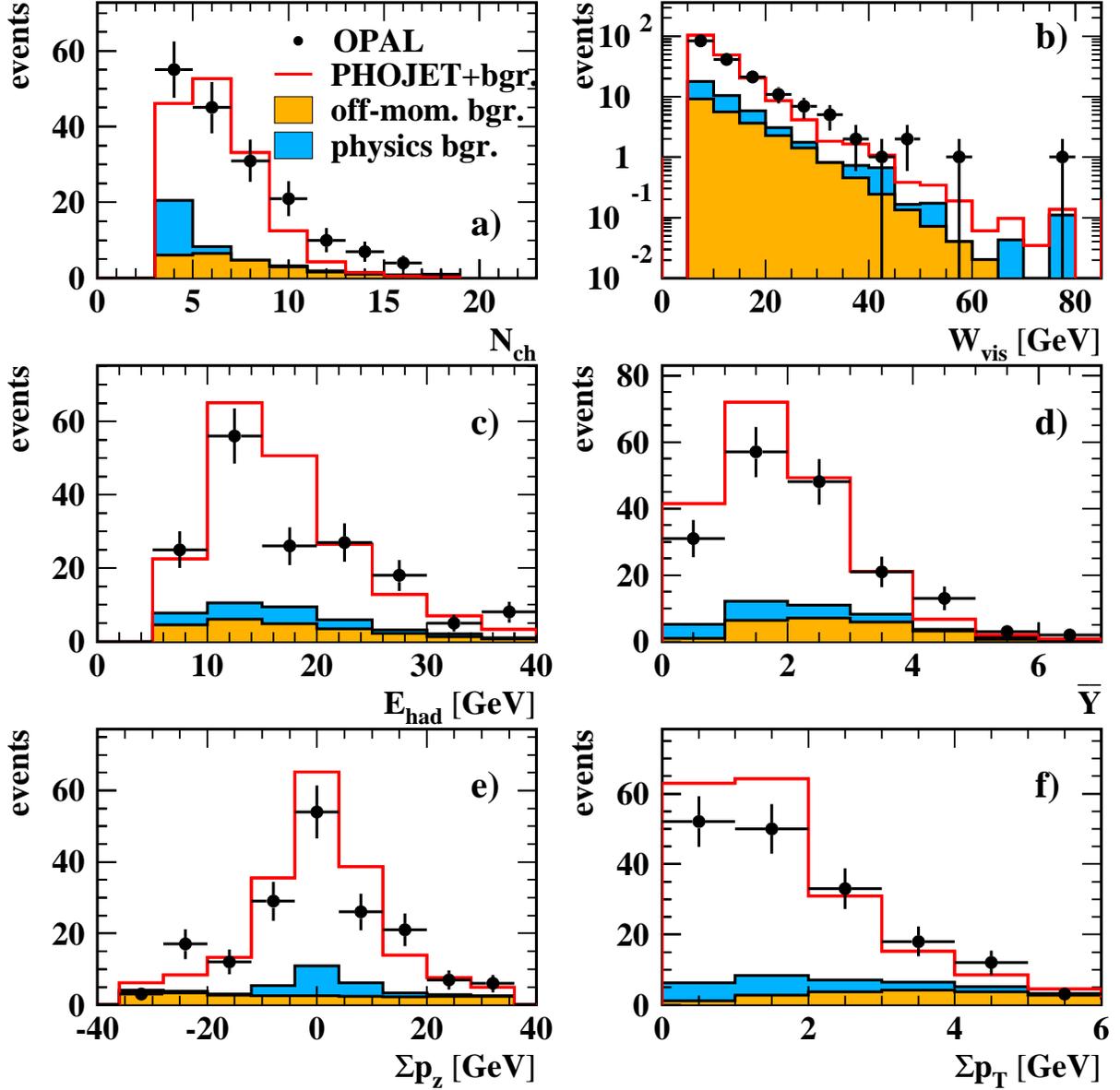}}
\caption{Distributions of 
(a) the number of  tracks of the hadronic final state,
(b) the visible hadronic invariant mass, (c) the total hadronic energy 
(d) the variable $\Ybar$, (e) the missing longitudinal momentum and
(f) the missing transverse momentum for double-tagged events.
The PHOJET1.10 predictions and backgrounds contributions are 
as in Fig.~\ref{comp_ele}.}
\label{comp_had}
\end{center}
\end{figure}

\begin{figure}[t]
\begin{center}
{\includegraphics[width=1.0\linewidth]{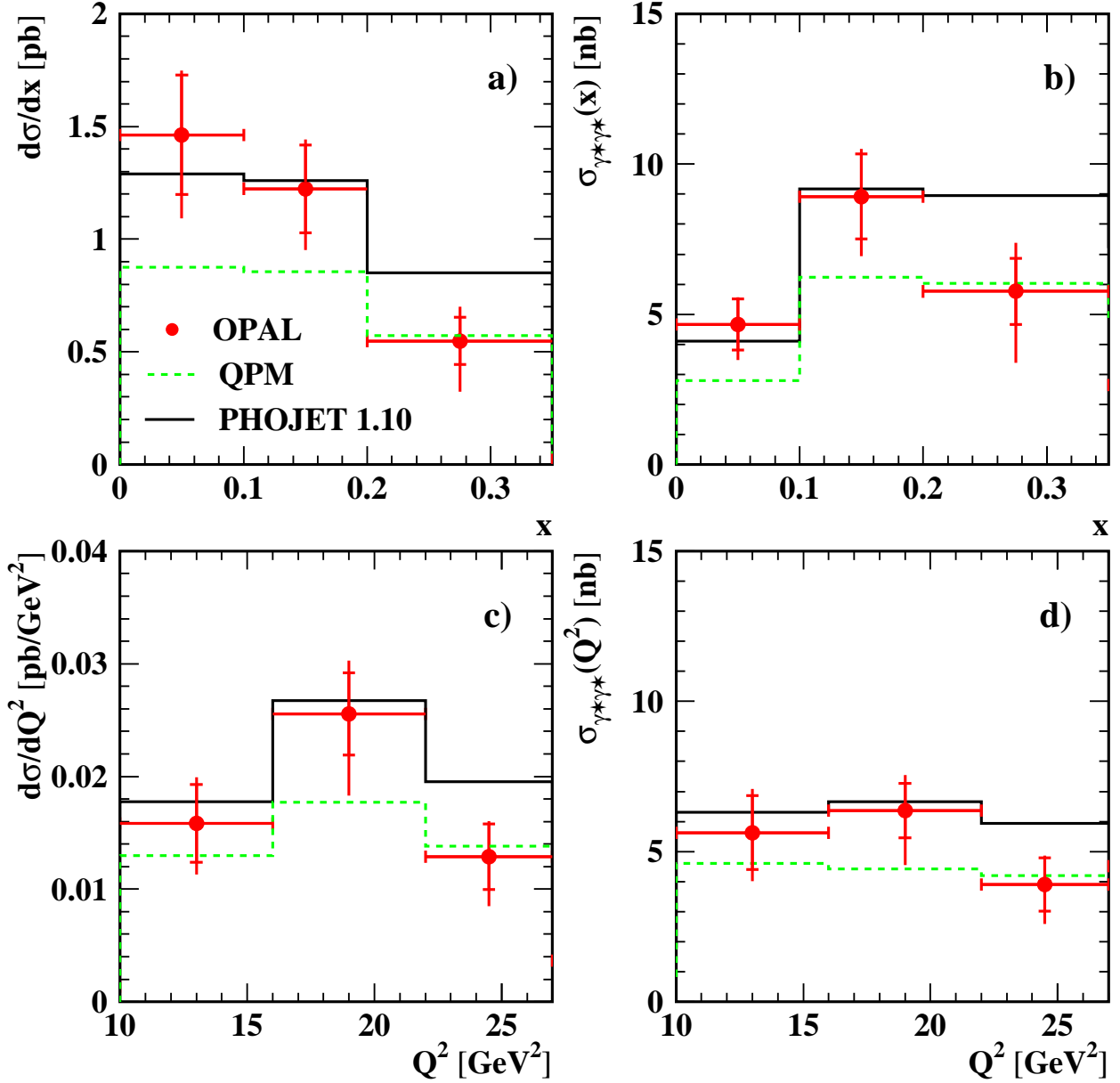}}
\caption{Cross-sections for the process 
$\epem \rightarrow  \epem\,\mbox{hadrons}$ 
in the region $E_{1,2}>0.4\eb$, $34<\theta_{1,2}<55$ mrad and $W>5$ GeV, 
and for the process $\ggss \rightarrow\,\mbox{hadrons}$,
as functions of $x$ for $\langle Q^2 \rangle = 17.9 $ GeV$^2$ (a,b), and
as functions of $Q^2$ (c,d).
Data are shown as full dots in the centre of the bins.
The inner error bars represent the statistical errors and the outer error
bars represent statistical and systematic errors added in quadrature.
The predictions of PHOJET1.10 are shown as  solid lines, and those of QPM as
dashed lines.}
\label{cros11}
\end{center}
\end{figure}

\begin{figure}[t]
\begin{center}
{\includegraphics[width=0.98\linewidth]{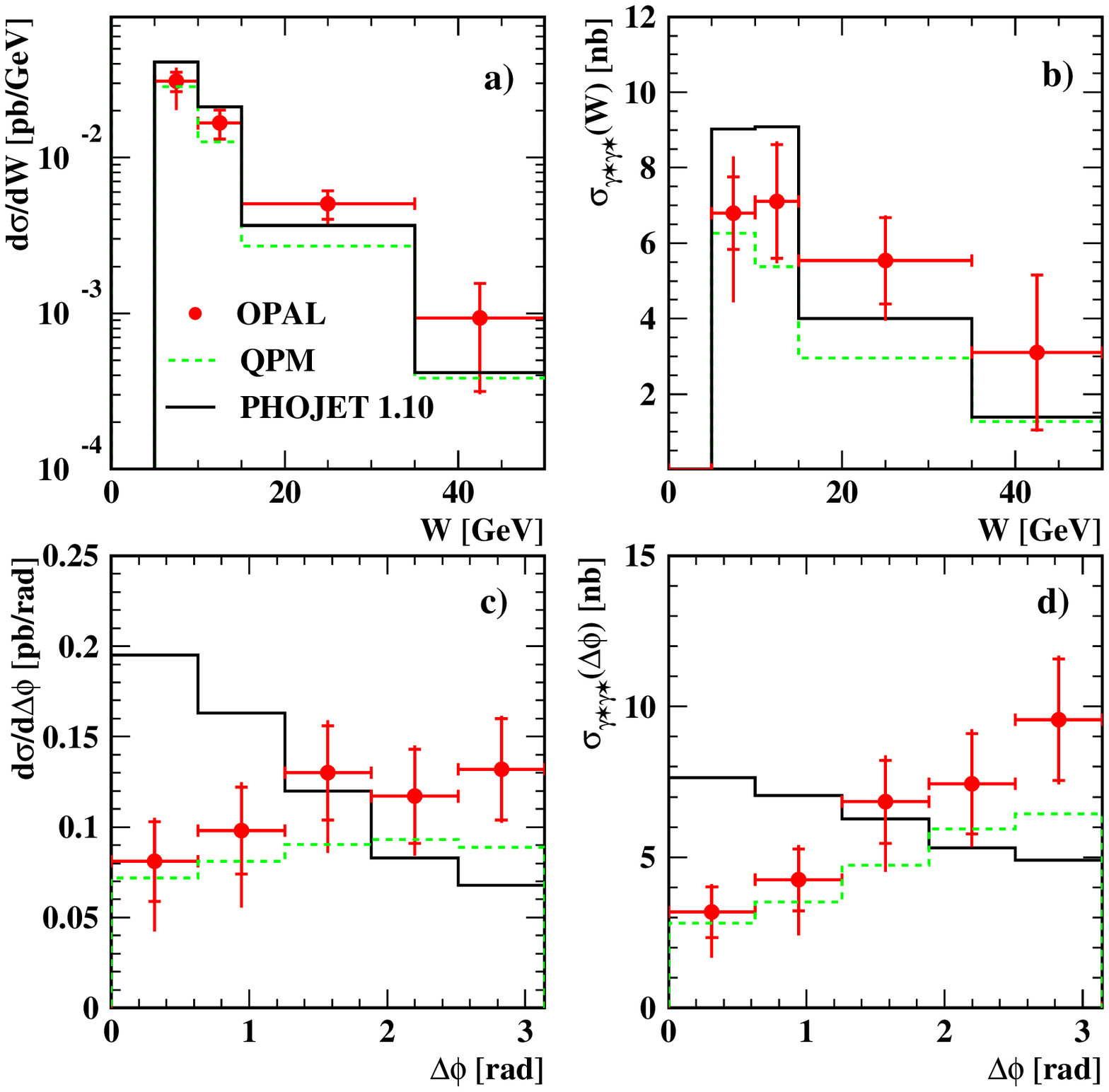}}
\caption{Cross-sections for the process 
         $\epem \rightarrow  \epem\,\mbox{hadrons}$ in the region 
         $E_{1,2}>0.4\eb$, $34<\theta_{1,2}<55$ mrad and $W>5$ GeV, 
         and for the process $\ggss \rightarrow\,\mbox{hadrons} $ 
         for $\langle Q^2\rangle = 17.9 $ GeV$^2$, 
         as  functions of $W$ (a,b) and $\Delta \phi$ (c,d). 
         Data are shown as full dots in the centre of the bins.
         The inner error bars represent the statistical errors and the
         outer error bars represent statistical and systematic errors
         added in quadrature. The predictions of PHOJET1.10 
         are shown as solid lines, and those of QPM as dashed lines.}
\label{cros12}
\end{center}
\end{figure}

\begin{figure}[t]
\begin{center}
{\includegraphics[width=0.98\linewidth]{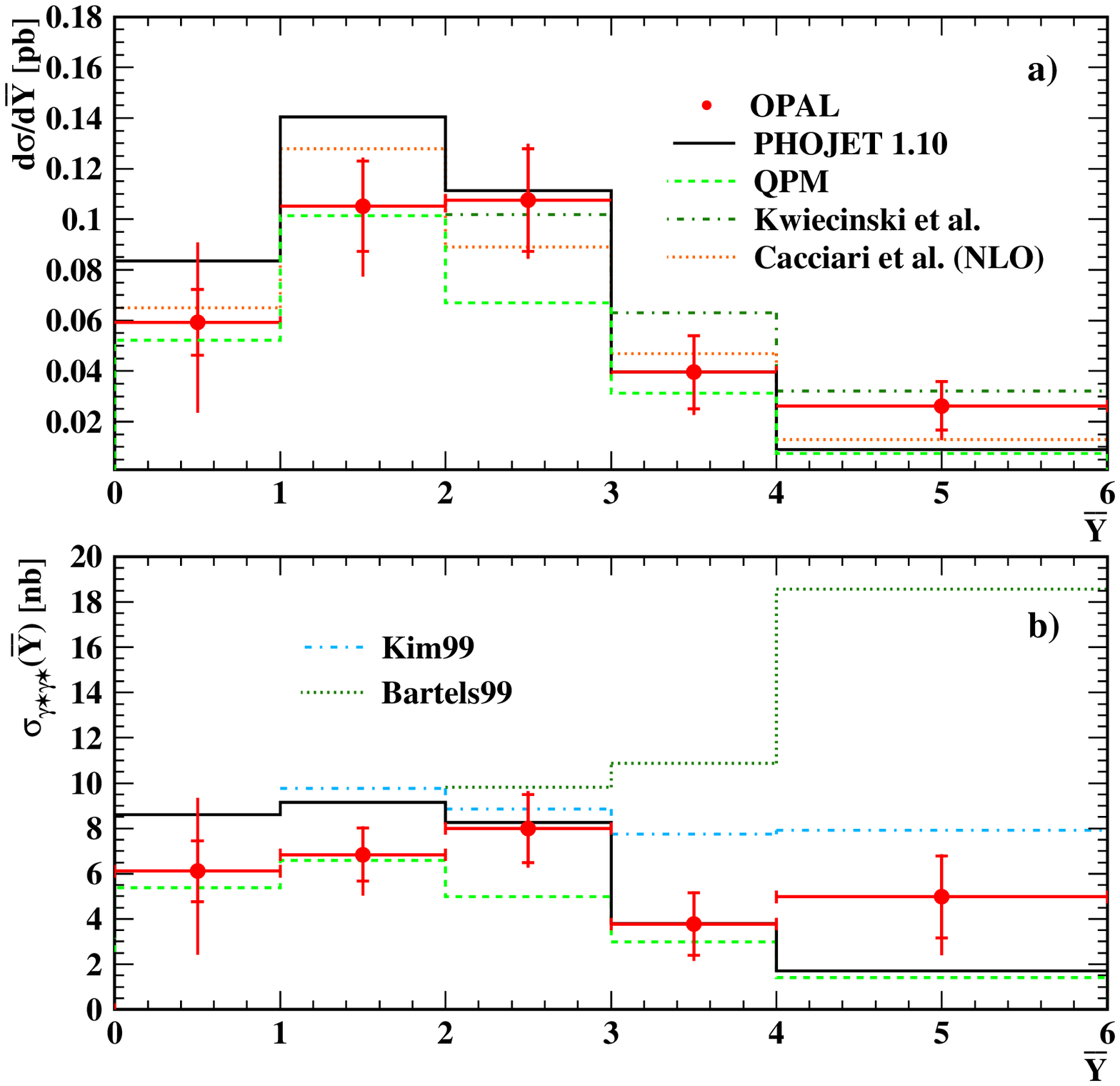}}
\caption{
Cross-sections for the process $\epem\rightarrow \epem\,\mbox{hadrons}$ 
in the region $E_{1,2}>0.4\eb$, $34<\theta_{1,2}<55$ mrad and $W>5$ GeV, 
and the process $\ggss \rightarrow\,\mbox{hadrons} $ 
for $\langle Q^2 \rangle = 17.9 $ GeV$^2$, as functions of $\Ybar$. Data are
shown as full dots in the centre of the bins. The inner error bars
represent the statistical errors and the outer error bars represent
statistical and systematic errors added in quadrature. The predictions
of PHOJET1.10 are shown as the solid lines, that of the NLO calculation of
the process \eeqq as dotted lines, and those of QPM as dashed lines.
Three BFKL calculations are shown: a LO one from Bartels et~al. (Bartels99),
NLO from Kim et~al. (Kim99) using $Y$, and the calculation from Kwiecinski
et~al., using the consistency constraint calculated for \Ybar. 
}
\label{cros13}
\end{center}
\end{figure}
\end{document}